\documentclass[useAMS,usenatbib]{mn2e}

\usepackage{lscape}
\usepackage{graphicx}
\usepackage{natbib}
\usepackage{graphics}
\usepackage{rotating,multirow}
\usepackage{subfigure}
\usepackage{colortbl}

\newcommand{\degree}{\hbox{$^\circ$}}

\def \vhel{\ifmmode{~V_{{\rm HEL}}}\else{~$V_{{\rm HEL}}$}\fi}
\def \vsys{\ifmmode{~V_{{\rm SYS}}}\else{~$V_{{\rm SYS}}$}\fi}
\def \HA {\ifmmode{{\rm\H}\alpha}\else{${\rm\ H}\alpha$}\fi}

\def \msun{\ifmmode{{\rm\ M}_\odot}\else{${\rm\ M}_\odot$}\fi}

\def \myr{\ifmmode{{\rm\ M}_\odot{\rm\ yr}^{-1}}
        \else{${\rm\ M}_\odot$ yr$^{-1}$}\fi}

\def \mdot{\ifmmode{\dot{M}}\else{$\dot{M}$}\fi}
\def \tena#1 #2 {\ifmmode{#1 \times 10^{#2}}\else{$#1 \times 10^{#2}$}\fi}
\def \kms{\ifmmode{~{\rm km\,s}^{-1}}\else{~km s$^{-1}$}\fi}

\title[High-resolution radio observations of GRS\,1915+105]{A decade of high-resolution radio observations of GRS\,1915+105}

\author[A. Rushton et al.]{A. Rushton$^1$\thanks{E-mail: Anthony.Rushton (`at') Manchester.ac.uk (AR)}, R. E. Spencer$^1$, G. Pooley$^2$ and S. Trushkin$^3$\\
\\
$^1$Jodrell Bank Centre for Astrophysics, School of Physics and Astronomy, The University of Manchester, Manchester M13 9PL\\
$^2$The University of Cambridge, Mullard Radio Astronomy Observatory, Cavendish Laboratory, J. J. Thomson Avenue, Cambridge CB3 0HE\\
$^3$Special Astrophysical Observatory RAS, Nizhnij Arkhyz, 369167, Russia
}

\begin{document}

\date{Accepted XXXX XXX XX. Received XXXX XXX XX; in original form XXXX XX XX}

\pagerange{\pageref{firstpage}--\pageref{lastpage}} \pubyear{2008}

\maketitle

\label{firstpage}

\begin{abstract}
The radio emitting X-ray binary GRS\,1915+105 shows a wide variety of X-ray and radio states. We present a decade of monitoring observations, with the RXTE-ASM and the Ryle Telescope, in conjunction with high-resolution radio observations using MERLIN and the VLBA. Linear polarisation at 1.4 and 1.6~GHz has been spatially resolved in the radio jets, on a scale of $\sim150$~mas and at flux densities of a few mJy. Depolarisation of the core occurs during radio flaring, associated with the ejection of relativistic knots of emission. We have identified the ejection at four epochs of X-ray flaring. Assuming no deceleration, proper motions of 16.5 to 27~mas per day have been observed, supporting the hypothesis of a varying angle to the line-of-sight per ejection, perhaps in a precessing jet.
\end{abstract}


\section{Introduction}

GRS\,1915+105 has proved to be one of the best laboratories for the study of relativistic physical environments, due to its high-brightness, periodic outbursts of superluminal ejecta and relative proximity. Since the first detection in August 1992 with the WATCH instrument on board the X-ray telescope GRANAT~\citep{1992IAUC.5590....2C}, GRS\,1915+105 has demonstrated some of the most spectacular high-energy physics within the Galaxy. Shortly after its discovery, a radio counterpart was discovered by \cite{1993IAUC.5773....2M} and later an infrared component was also identified \citep{1993IAUC.5830....1M}. Its radio light-curve shows a highly variable and complex structure, with spatially resolved features that can be followed over many days.

A clear correlation between the X-ray, infrared and radio emission was quickly established by~\cite{1994Natur.371...46M}, with rapid time variability in each band. Radio flaring was found to correspond to a rapid change in the hard X-rays and possible production of high-energy gamma rays. In March 1994, 20~cm VLA observations of GRS\,1915+105 detected the first superluminal motion of a Galactic source \citep{1994Natur.371...46M}. This major breakthrough provided the direct evidence of relativistic jets and an extreme physical environment within the Galaxy. The name ``microquasar'' was coined due to their obvious similarities with their extra-galactic counterparts, quasars.

The launch of the Rossi X-ray Timing Explorer (RXTE) satellite, in December 1995, signified the start of a long-term monitoring campaign of X-ray binaries. The All Sky Monitor (ASM) on board the RXTE has taken daily observations of GRS\,1915+105 since its launch. \cite{1996ApJ...473L.107G} detected unusual X-ray variability on time scales of under one second to days. The RXTE-ASM lightcurve was found to be both highly complex and structured, due to instabilities in the accretion disc. Detailed observations with the RXTE's Proportional Counter Array (PCA) instrument (\citealt{1997ApJ...477L..41C}; \citealt{1997ApJ...488L.109B}) identified different spectral states, including a low-hard state dominated by a power-law and a high-soft state with a strong disc-blackbody component. A transition between such states is believed to be associated with the ejection of superluminal plasmons \citep{1999MNRAS.304..865F}.

\begin{table*}

\begin{center}\begin{tabular}{|>{\columncolor[rgb]{0.8,0.8,0.8}}l|ccccccc|}
\hline
\rowcolor[rgb]{0.8,0.8,0.8} Epoch (MJD) & Frequency/Array  & $I_{\rm{total}}$ &  $LP_{\rm{total}}$ &   Fraction & Noise & Spectral$^{\ddagger}$ & Resolved? \\
\rowcolor[rgb]{0.8,0.8,0.8}      & (GHz) &  (mJy)       &   (mJy) &   (\%)            & ($\mu$Jy/beam)    &  Index  &      \\

\hline 
$^{\star}$1997 October 31 (50752) -core & 2.27 / VLBA &  $105\pm2$ &    $\le0.1^{\dagger}$ &  $\le0.5^{\dagger}$ & 430 & \multirow{4}{*}{-} & \multirow{4}{*}{YES} \\
$^{\star}$1997 October 31 (50752) -jet & 2.27 / VLBA & $226\pm2$ &  $5.6\pm0.3$ & 2.5 & 430 & & \\
$^{\star}$1997 October 31 (50752) -core & 8.3 / VLBA  & $33\pm1$ &    $\le0.1^{\dagger}$& $\le0.5^{\dagger}$ & 460 & &\\
$^{\star}$1997 October 31 (50752) -jet & 8.3 / VLBA  & $65\pm3$ & $4.7\pm0.3$& 7.2  & 460 & &\\
\hline
1998 June 9 (50973) & 1.6 / MERLIN &  $145.1\pm0.4$ &  $9.5\pm0.3$&  $6.9\pm0.2$ & 300  & -0.3 & NO\\
\hline
1999 April 2 (51270) & 1.6 / MERLIN &  238.4$\pm0.3$ &  $3.5\pm0.3$& $1.5\pm0.2$ & 530  & -0.6& NO\\
\hline
1999 November 18 (51500) & 1.6 / MERLIN &  221.3$\pm0.8$ &  $6.9\pm0.3$ & $3.2\pm0.2$  & 860 & -0.4& NO\\
\hline
1999 November 22 (51504) & 1.6 / MERLIN &  37.1$\pm0.7$ &  $\le0.1^{\dagger}$ & $\le0.4^{\dagger}$ & 300 & -0.2& NO\\
\hline
1999 December 28 (51540) & 1.6 / MERLIN &  89.1$\pm0.7$ & $12.7\pm1.6$  & $14.1\pm0.2$ & 400 & -0.4& NO\\
\hline
2003 March 6 (52704) & 1.6 / MERLIN &  $31.3\pm0.3$ &  $1.6\pm0.3$ & $5.2\pm1.0$ & 240 & -0.4& NO\\ 
\hline
2003 March 25 (52723) & 1.6 / MERLIN & $125.8\pm0.1$ &  $7.1\pm0.2$ & $5.6\pm0.2$  & 140 & 0.2& NO\\
\hline
2003 April 18 (52747) -core & 1.6 / MERLIN &  $135.4\pm0.4$ &  $3.5\pm0.2$ & $2.6\pm0.2$  & 270 & \multirow{2}{*}{0.0}& \multirow{2}{*}{YES}\\
2003 April 18 (52747) -jet & 1.6 / MERLIN &  $20.0\pm0.3$ &  $1.3\pm0.1$ & $6.5\pm0.5$  & 270 & &\\
\hline
2003 May 09 (52768) & 1.6 / MERLIN &  $43.0\pm0.1$ &  $2.7\pm0.1$ & $6.3\pm0.2$  & 160 & 0.2&NO\\
\hline
2003 June 15 (52805) -core & 1.6 / MERLIN &  $41.4\pm0.1$ &  $\le0.1^{\dagger}$  & $\le0.2^{\dagger}$ & 390 & \multirow{2}{*}{-0.4}& \multirow{2}{*}{YES}\\
2003 June 15 (52805) -jet & 1.6 / MERLIN &  $67.5\pm0.1$ &  $7.6\pm0.4$  & $11.9\pm0.6$ & 390 & & \\ 
\hline
2006 December 24 (54093) & 1.4 / MERLIN &  $145.0\pm0.5$ &  $7.7\pm0.6$  & $5.3\pm0.4$ & 380 & \multirow{2}{*}{-0.1}& \multirow{2}{*}{NO}\\
2006 December 24 (54093) & 1.6 / MERLIN &  $156.9\pm1.2$ &  $11.5\pm0.5$ &  $7.3\pm0.5$ & 380 && \\
\hline
2006 December 27 (54096) & 1.4 / MERLIN &  $20.7\pm0.5$ &  $\le0.1^{\dagger}$  & $\le0.5^{\dagger}$ & 230 & \multirow{2}{*}{-0.4} & \multirow{2}{*}{NO}\\
2006 December 27 (54096) & 1.6 / MERLIN &  $27.0\pm0.5$ &  $\le0.1^{\dagger}$  & $\le0.5^{\dagger}$ & 230 & &\\
\hline
2006 December 28 (54097) & 1.4 / MERLIN &  $25.8\pm0.4$ &  $\le0.1^{\dagger}$  & $\le0.4^{\dagger}$ & 210 & \multirow{2}{*}{-}& \multirow{2}{*}{NO}\\
2006 December 28 (54097) & 1.6 / MERLIN &  $26.2\pm0.4$ &  $\le0.1^{\dagger}$  & $\le0.4^{\dagger}$ & 210 & &\\
\hline
2007 January 04 (54104) & 1.4 / MERLIN &  $21.0\pm0.5$ &  $\le0.1^{\dagger}$ & $\le0.5^{\dagger}$ & 260 & \multirow{2}{*}{-0.3}& \multirow{2}{*}{NO}\\
2007 January 04 (54104) & 1.6 / MERLIN &  $22.2\pm0.5$ &  $\le0.1^{\dagger}$  & $\le0.5^{\dagger}$ & 260 & &\\
\hline 
\end{tabular}\end{center}
\caption{\label{table:journal}Journal of observations and results: Date of observation, observed frequency and array, total flux, total linearly polarised flux, fraction of total flux that is linearly polarised, systematic noise, spectral index and detection of resolved structure. [Notes: $^{\star}$ The observations were first published by \citealt{2000ApJ...543..373D}; $^{\dagger}$ Estimates are based on a upper limit of detection; $^{\ddagger}$ The spectral index is derived from the total MERLIN flux (i.e. core + jet) and the 15 GHz RT data, where $S \propto \nu^\alpha$.]}
\end{table*}

Radio polarisation was first detected with MERLIN in GRS\,1915+105 by \cite{1999MNRAS.304..865F} in the first four epochs (i.e. first four days) of the October 1997 flare at a frequency of 5~GHz. A clear asymmetry in the location of the polarised emission was found, showing the strongest detection of linearly polarised emission within the approaching jet, weaker in the receding jet and no detection ($<3\%$ of the total flux density) within the core. The polarisation position angle (PA) was found to rotate by at least $75^\circ$ over the four days, leading to the suggestion that the changes in polarisation could be due to Faraday effects (i.e. changes in the Faraday depth as the plasmons move away from the core), implying a change in rotation measure of $\Delta RM>300$~rad~m$^{-2}$.

This paper describes observations taken over a decade with MERLIN at 18 and 21~cm. The observations were either in conjunction with INTEGRAL observations as part of the Galactic Plane Survey~\citep{2004ApJ...607L..33B,2004ESASP.552..321F}, or were triggered by flare events, found by other radio telescopes. Polarisation behaviour, structural variations and the relationship of activity in the radio regime to the X-ray behaviour were investigated.

\section{Observations and data reduction}

\subsection{RXTE All Sky Monitor}

The All Sky Monitor (ASM) instrument on-board the RXTE has been monitoring the sky since March 1996 and the data presented here covers the period from March 1997 to 2007. With each orbit of the RXTE, the ASM surveyed $\sim80\%$ of the sky to a depth of $20-100$~mCrab, making approximately ten observations of a source per day. A more detailed description of the RXTE-ASM can be found in~\cite{1996ApJ...469L..33L}.

Each individual pointing, or dwell, was a 90 second integration of the source, with intensities measured in three energy bands of $1.5-3$, $3-5$, and $5-12$~keV. The Crab Nebula flux between $1.5-12$~keV corresponds to about 75 ASM counts~s$^{-1}$. To calculate the spectral hardness (HR2), individual dwells were averaged into daily points and the ratios between the $5-12$~keV and $1.5-3$~keV energy bands were taken. 

\subsection{Ryle Telescope 15 GHz monitoring}

The Ryle Telescope (RT) is a linear east-west radio interferometer located at the Mullard Radio Astronomy Observatory, UK. The array operates at a frequency of 15~GHz with associated baselines between approximately 18~metres and 4.8~kilometres.

An extensive monitoring campaign began with the RT in 1996 of a few radio-bright X-ray binaries (including GRS\,1915+105), which coincided with the launch of the RXTE satellite. Observations of target sources were interleaved with a nearby phase calibrator (B1920+154) and the flux-density scale was set by short scans of 3C~48 and 3C~286. The data were sampled every eight seconds and averaged into five minute data points with an RMS of $\sim2$~mJy.

\begin{figure*}
  \begin{center}
\includegraphics[width=16cm, angle=0, trim=0 -10 0 0]{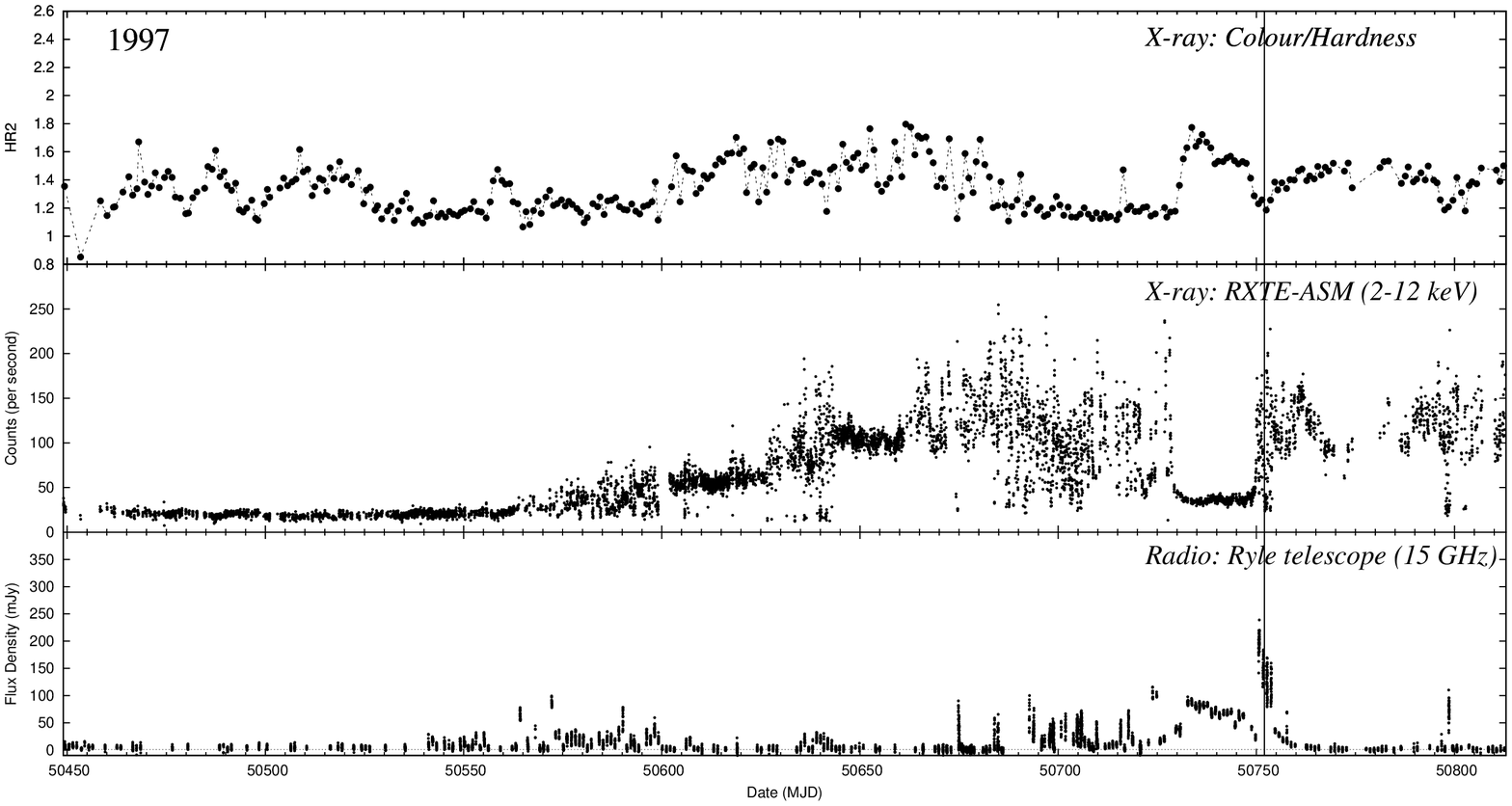}
\includegraphics[width=16cm, angle=0, trim=0 -10 0 0]{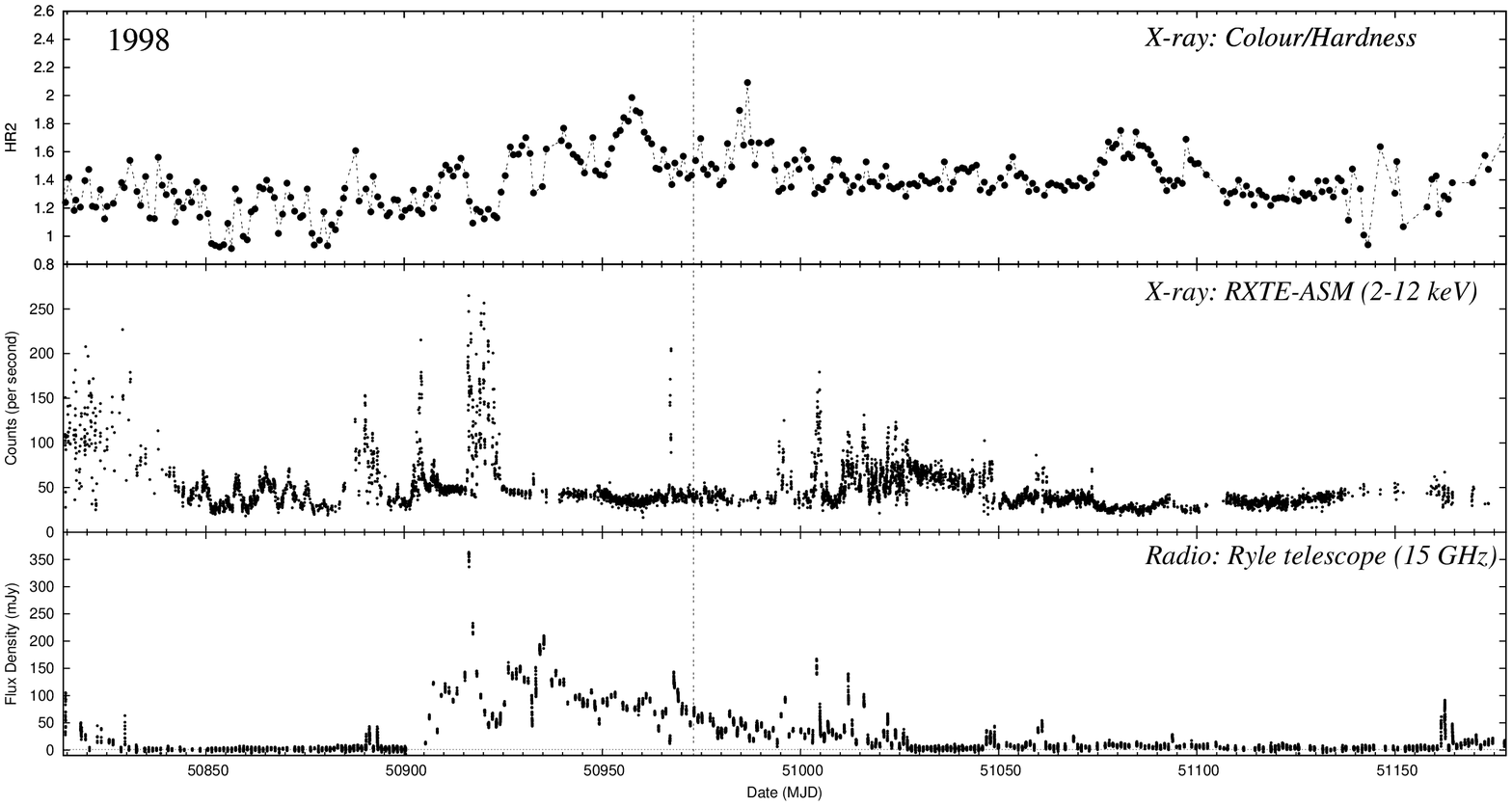}
\end{center}
\caption[RM]{\label{monitor_1997-1998}X-ray observations with the All Sky Monitoring instrument on board the RXTE satellite and 15~GHz radio monitoring with the Ryle Telescope in 1997 and 1998. The dotted line denotes a MERLIN 18 cm epoch (1998 June 9), showing that the source was in the `plateau-state'. [N.B. The solid line in 1997 marks a VLBA observation (1997 October 31) originally published by \cite{2000ApJ...543..373D}].}
\end{figure*}

\begin{figure*}
\begin{center}
\includegraphics[width=16cm, angle=0, trim=0 -10 0 0]{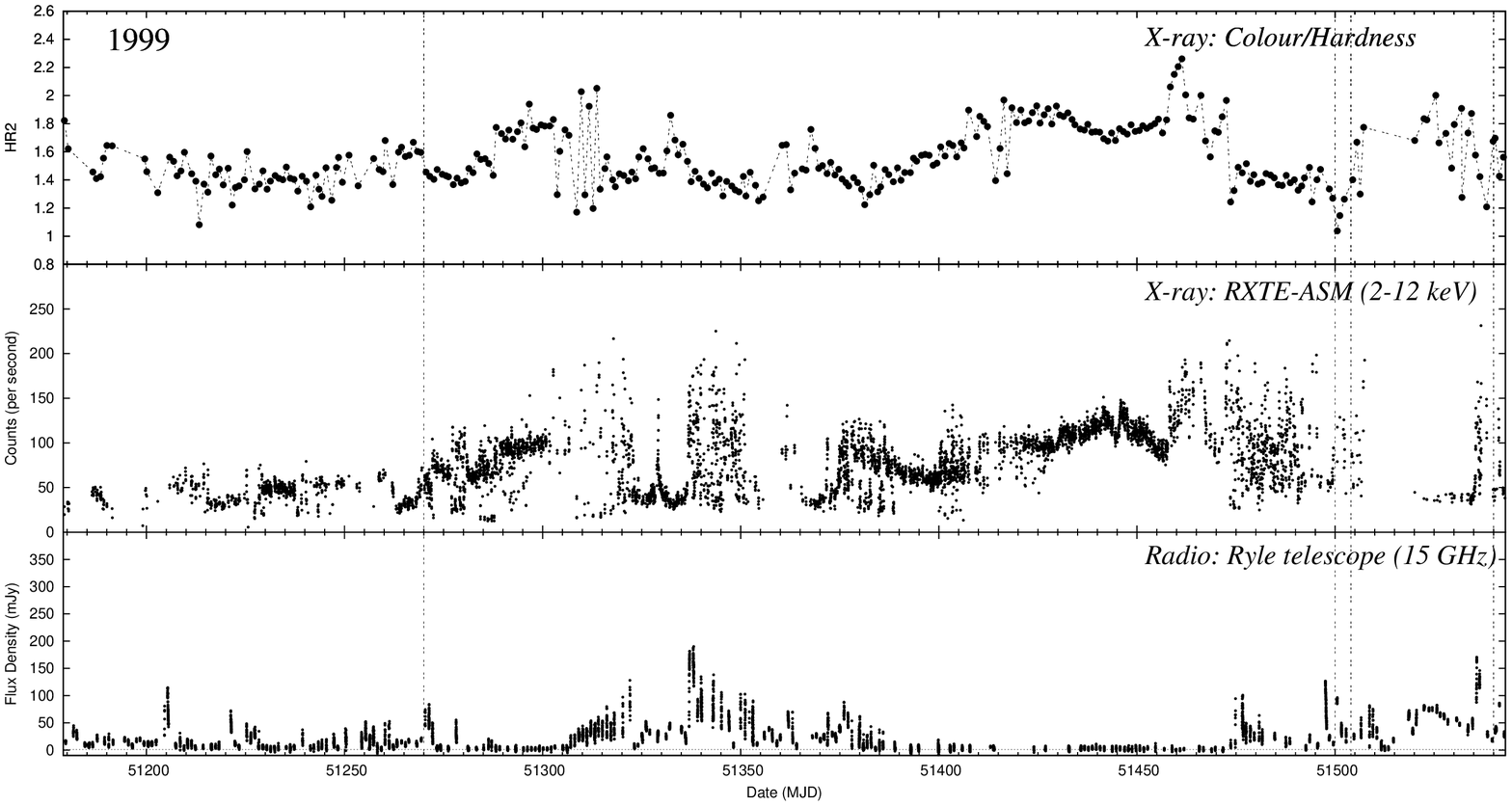}
\includegraphics[width=16cm, angle=0, trim=0 -10 0 0]{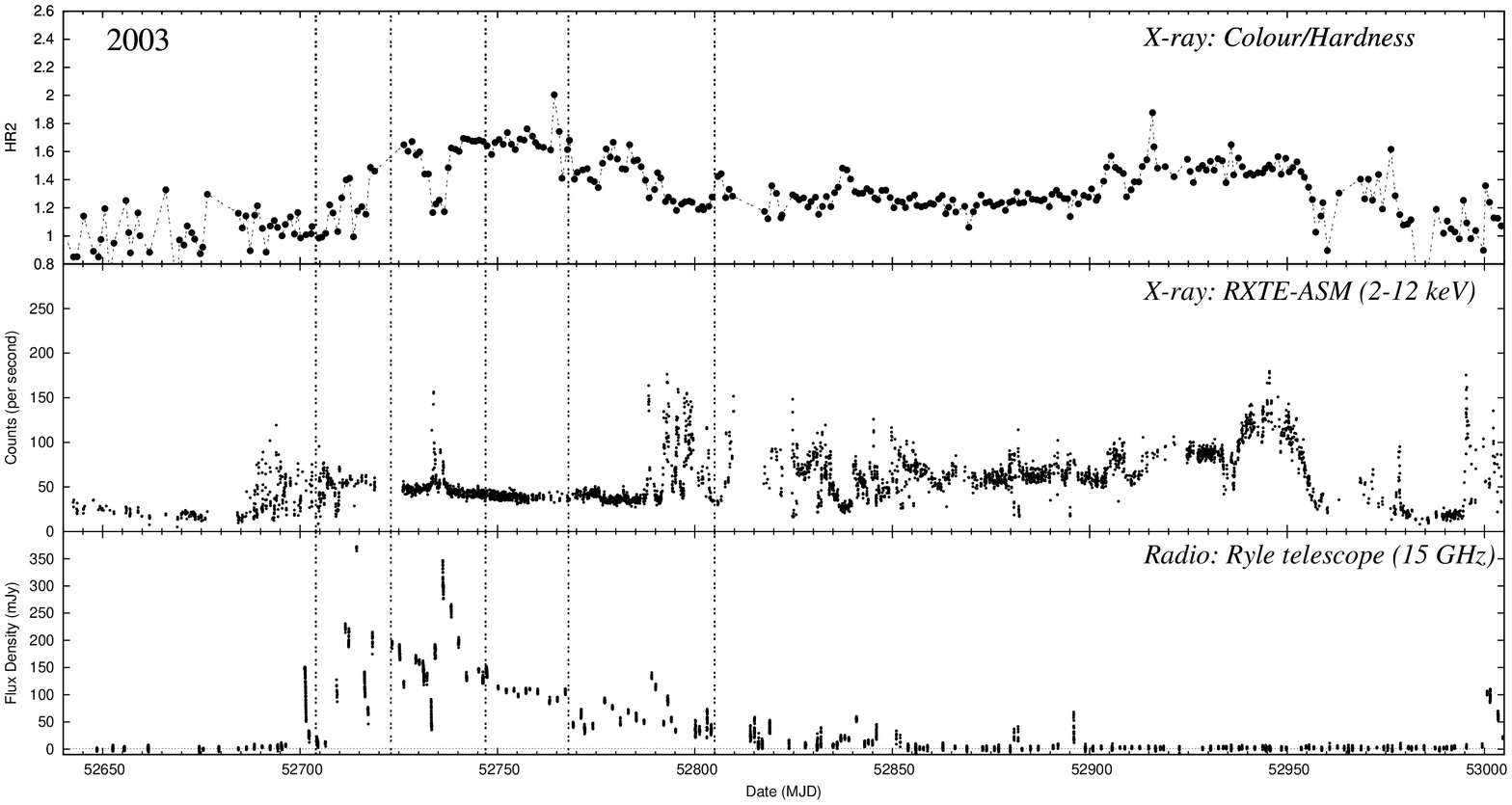}
\includegraphics[width=16cm, angle=0, trim=0 -10 0 0]{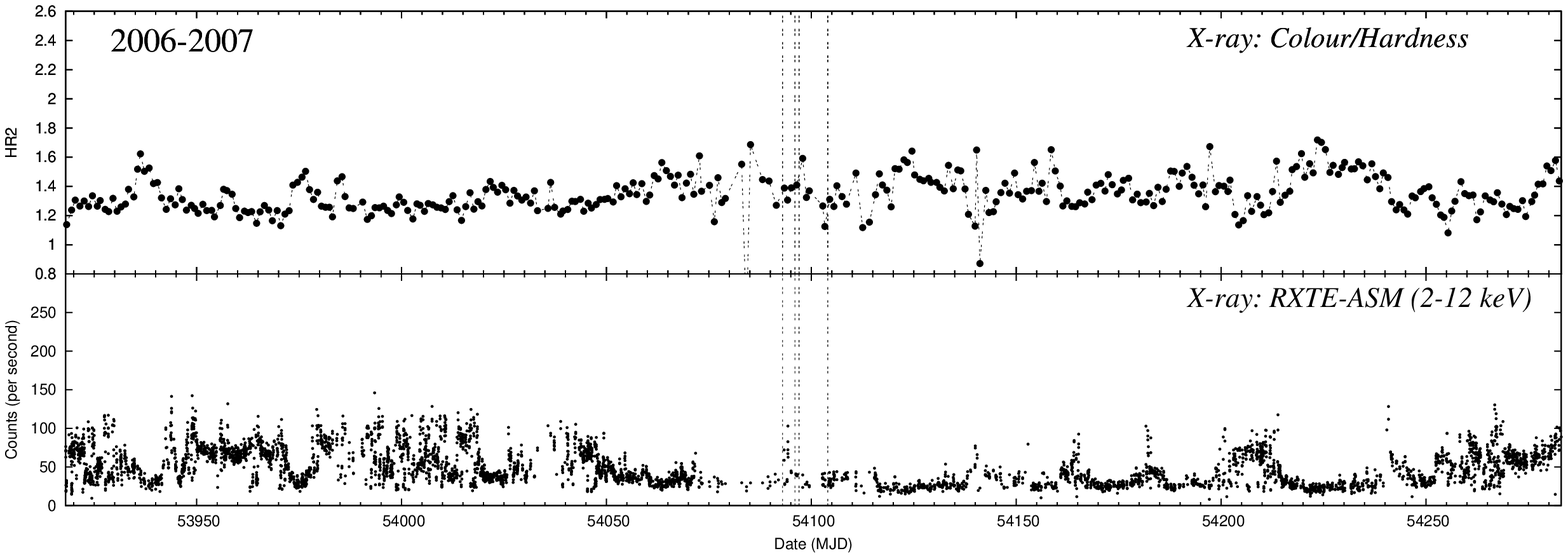}
\end{center}
\caption[RM]{\label{monitor_1999-2003-2006}X-ray observations with the All Sky Monitoring instrument on board the RXTE satellite and 15~GHz radio monitoring with the Ryle Telescope in 1999, 2003 and 2006 -- 2007 (no RT data were available in 2006 -- 2007). The dotted lines denote the MERLIN 18~cm epochs, showing that the source was in different X-ray `states' during the observations.}
\end{figure*}

\cite{1997MNRAS.292..925P} describe the details of this programme, indicating the detection of a 20 -- 40 minute quasi-periodic variation of GRS\,1915+105 at a frequency of 15 GHz, possibly coupled with variations in the soft X-rays. The results presented in Section~\ref{MERLIN_results} were taken during monitoring periods with the RT and RXTE-ASM. Figures~\ref{monitor_1997-1998} and \ref{monitor_1999-2003-2006} show data collected during 1997, 1998, 1999, 2003 and 2006 -- 2007, each for a period of 365 days (note RT data was not available in 2006 -- 2007).

\subsection{RATAN-600}

The RATAN-600 radio telescope consists of a 576 metre circle of radio reflectors, located  at the Special Astrophysical Observatory, Russia. The telescope can operate at multiple radio frequencies between 610 MHz and 30 GHz.

RATAN-600 observations were carried out as part of a long-term monitoring programme on variable X-ray sources (\citealt{2000A&AT...19..525T};  \citealt{2006smqw.confE..15T}). Observations of GRS\`1915+105 were taken at 1.0, 2.3, 4.8, 7.7 and 11.2~GHz during the 2006 -- 2007 flare: on 2006 December 22 (MJD~54091), 24 (MJD~54093), 27 (MJD~54096) and 2007 January 04 (MJD~54104). A standard continuum radio system with cryogenic receivers made observations at 4.8, 7.7 and 11.2~GHz. At lower frequencies, a low-noise HEMT-based radiometer operated at 1 and 2.3~GHz. The observations were carried out with the `Northern sector' antenna of RATAN-600\footnote{http://w0.sao.ru/ratan}.

\subsection{VLBA}

The Very Long Baseline Array (VLBA) is a network of remotely controlled radio telescopes distributed across the USA. The Maximum baselines are over 8500 kilometres producing a sub-milliarcsecond resolution at centimetre wavelengths.

Observations taken on 1997 October 31 (MJD~50752) with the VLBA have been presented by~\cite{2000ApJ...543..373D}. The VLBA data have been reprocessed in order to study linearly polarised emission from the source. The observations were originally triggered by daily monitoring with the Green Bank Interferometer (GBI) during a period of radio flaring. It was possible to obtain data from multiple frequencies by means of a frequency-selective optics allowing the simultaneous use of 3.6 and 13~cm (8.4 and 2.27~GHz) receivers.

\subsection{MERLIN}

The Multi-Element Radio Linked Interferometer Network  (MERLIN) array connects a total of seven telescope across the UK: Mark II and Lovell at Jodrell Bank, the 32 metre at Cambridge and 25 metre antennas at Knocking, Darnhall, Pickmere and Defford (although the Lovell was not included in all observations). MERLIN observations of GRS\,1915+105 have been made during 14 epochs since 1998 (see Table~\ref{table:journal}) at 1.4 and 1.6~GHz, where the angular resolution was 150 and 130~mas, respectively. Note that MERLIN observations at 5~GHz have been described by~\cite{1999MNRAS.304..865F} and \cite{2007MNRAS.375.1087M}.

The antennas have orthogonal right (R) and left (L) circular polarisation feeds from which RR, LL, RL and LR cross-correlations were formed, measuring 15 MHz bandwidths correlated into 1 MHz channels. Data correlation was performed by the dedicated MERLIN correlator located at Jodrell Bank, UK. Preliminary flux calibration and gain elevation corrections were applied to the data using local MERLIN software and transformed into \textsc{fits} format. The data were loaded into NRAO's Astronomical Image Processing System (\textsc{aips}) and bad data were removed using the task \textsc{ibled}. The flux density scale was determined from observations of 3C~286, using the three short baselines between the Mark II, Pickmere and Darnhall telescopes. A flux density of 13.5~Jy at 1.6~GHz was assumed for 3C~286 \citep{1977A&A....61...99B}, from which we derived a flux density for the point source calibrator OQ~208. Scans of GRS\,1915+105 were interleaved with the phase reference source, 1919+086, with a 5:3 minute cycle time respectively, allowing the use of antenna amplitude and phase-gain solutions from the calibrator source using the tasks \textsc{calib} and \textsc{clcal}. 

The \textsc{aips} task \textsc{pcal} was used on 1919+086 to determine the effective feed polarisation parameters for each individual telescope and these were written into the antenna file (or AN table) for future correction. This corrects the leakage of flux from one cross-polarised mode to the other, assuming the phase and amplitude of the source has been correctly calculated and is flat across the band (i.e. phase referenced). The instrumental response due to changes in the parallactic angle was also removed using observations of the phase reference source 1919+086.

Next, any systematic phase delay between the right and left hands needed to be corrected by a source with a known polarisation angle. A phase delay difference between right and left hands was removed before performing \textsc{pcal}. Whilst applying the solutions from \textsc{pcal}, a rotation of Q + iU corrected the R-L phase delay was achieved using the \textsc{aips} task \textsc{clcor}. The known position angle of the polarization of 3C~286 was used to calibrate the phase differences between L and R for each antenna.

Finally the target source went through a few rounds of phase-only self-calibration. The calibrated \textit{uv} data of GRS\,1915+105 were then Fourier transformed and the \textsc{clean} algorithm was applied using the \textsc{aips} task \textsc{imagr}. Each image was made using approximately 12 hours of MERLIN data. Table~\ref{table:journal} shows the total flux density and the polarized flux density from the images. Also shown are the results of VLBA observations which were re-processed using data from~\cite{2000ApJ...543..373D}. These values are presented for a multi-wavelength comparison of the polarisation levels in the core and jet.

\begin{figure}
\includegraphics[width=8.5cm, trim=0 -10 0 0]{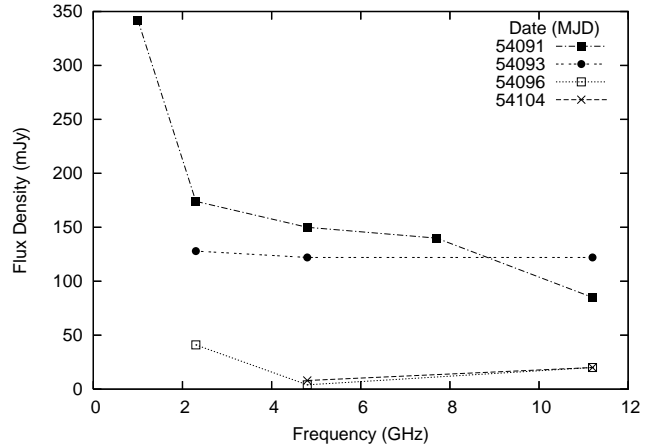}
\caption[RM]{\label{RATAN}RATAN observations of the 2006 -- 2007 flare at the frequencies 1.0, 2.3, 4.8, 7.7 and 11.2~GHz.}
\end{figure}

\begin{figure*}
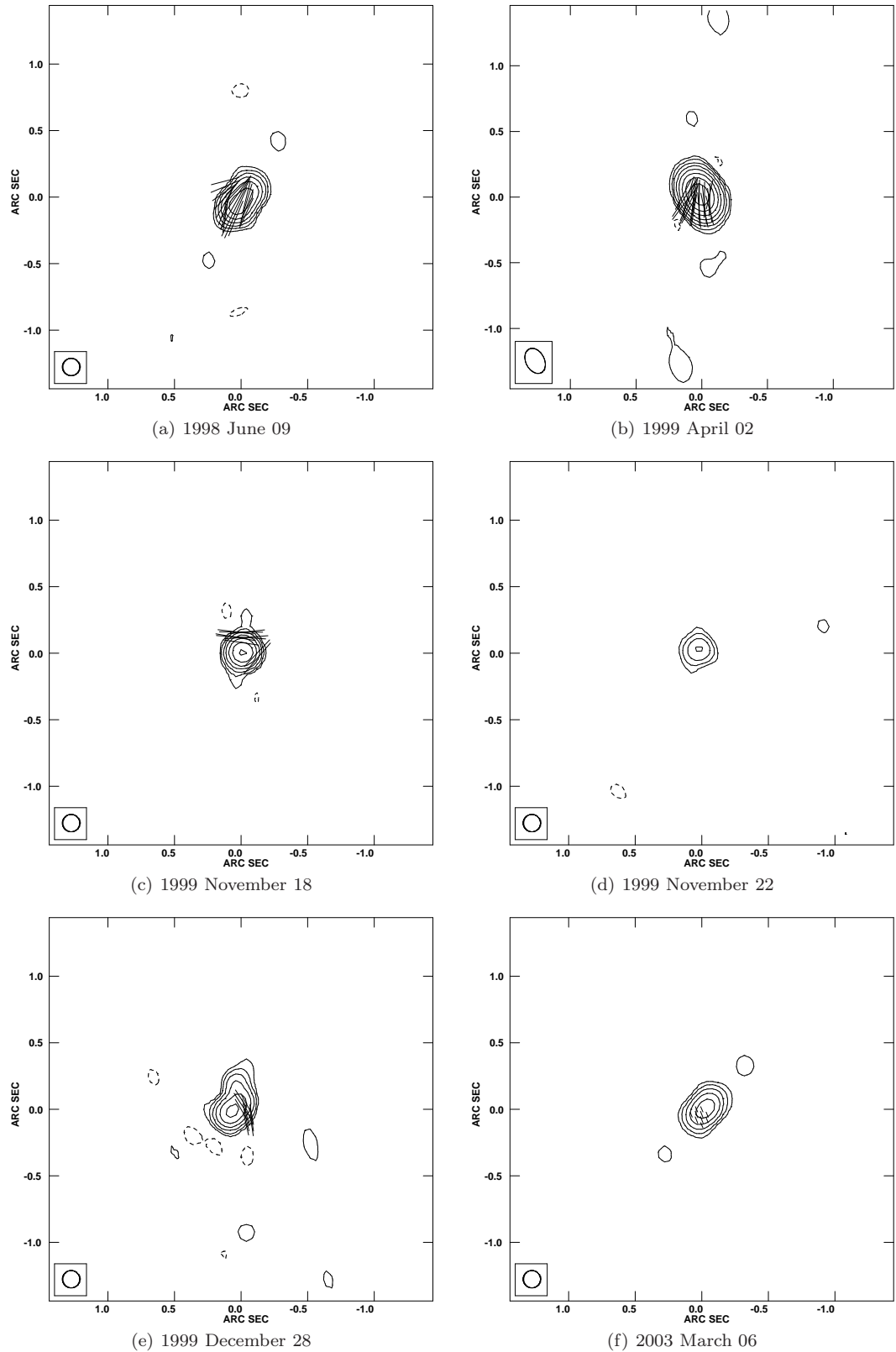

    \centering
    \subfigure[1998 June 09]
    {
\includegraphics[width=6.7cm, angle=-90]{98JUN03-nofooter.ps}
        \label{fig:first_sub}
    }
    \subfigure[1999 April 02]
    {
\includegraphics[width=6.7cm, angle=-90]{99APR02.ps}
        \label{fig:first_sub}
    }
    \subfigure[1999 November 18]
    {
\includegraphics[width=6.7cm, angle=-90]{18Nov1999.ps}
        \label{fig:first_sub}
    }
    \subfigure[1999 November 22]
    {
\includegraphics[width=6.7cm, angle=-90]{22Nov1999.ps}
        \label{fig:second_sub}
    }
    \subfigure[1999 December 28]
    {
 \includegraphics[width=6.7cm, angle=-90]{28dec1999.ps}
        \label{fig:third_sub}
    }
    \subfigure[2003 March 06]
    {
\includegraphics[width=6.7cm, angle=-90]{03MAR05-NOFOOTER.PS}
        \label{fig:second_sub}
    }
    \\
\caption[sticks]{\label{maps-A}MERLIN images of GRS\,1915+105 at~1.6 GHz (top left to bottom right) centred on Right Ascension~19~15~11.54801 Declination~+10~56~44.7662, on 1998 June 3, 1999 April 02, November 18, 22, December 28 and 2003 March 06. The restoring beam is shown in the lower corner of each image and the stick vectors represent the mean position angle of linear polarisation with an intensity of 12.5 mJy per arcsecond. The contours are $3\sigma\times-1, 1, 2, 4, 8, 16....n$. (a) - 1998 June 09. $3\sigma=$~0.90 mJy with a peak flux density = 95.1 mJy. (b) - 1999 April 02. $3\sigma=$~1.6 mJy with a peak flux density = 221.7 mJy. (c) - 1999 November 18.  $3\sigma=$~2.6 mJy with a peak flux density = 194.8~mJy. (d) - 1999 November 22.  $3\sigma=$~0.9 mJy with a peak flux density = 8.0~mJy. (e) - 1999 December 28.  $3\sigma=$~1.2 mJy with a peak flux density = 49.2~mJy. (f) - 2003 March 06.  $3\sigma=$~0.733 mJy with a peak flux density = 22.3~mJy.}
\end{figure*}

\begin{figure*}
    \centering
    \subfigure[2003 March 25]
    {
\includegraphics[width=6.7cm, angle=-90]{03MAR01-NOFOOTER.PS}
        \label{fig:third_sub}
    }
    \subfigure[2003 April 18]
    {
\includegraphics[width=6.7cm, angle=-90]{03APR17-NOFOOTER.PS}
        \label{fig:first_sub}
    }
    \subfigure[2003 May 09]
    {
\includegraphics[width=6.7cm, angle=-90]{image03May09.ps}
        \label{fig:first_sub}
    }
    \subfigure[2003 June 15]
    {
\includegraphics[width=6.7cm, angle=-90]{03JUN15-NOFOOTER.PS}
        \label{fig:second_sub}
    }
    \subfigure[2006 December 24]
    {
 \includegraphics[width=6.7cm, angle=-90]{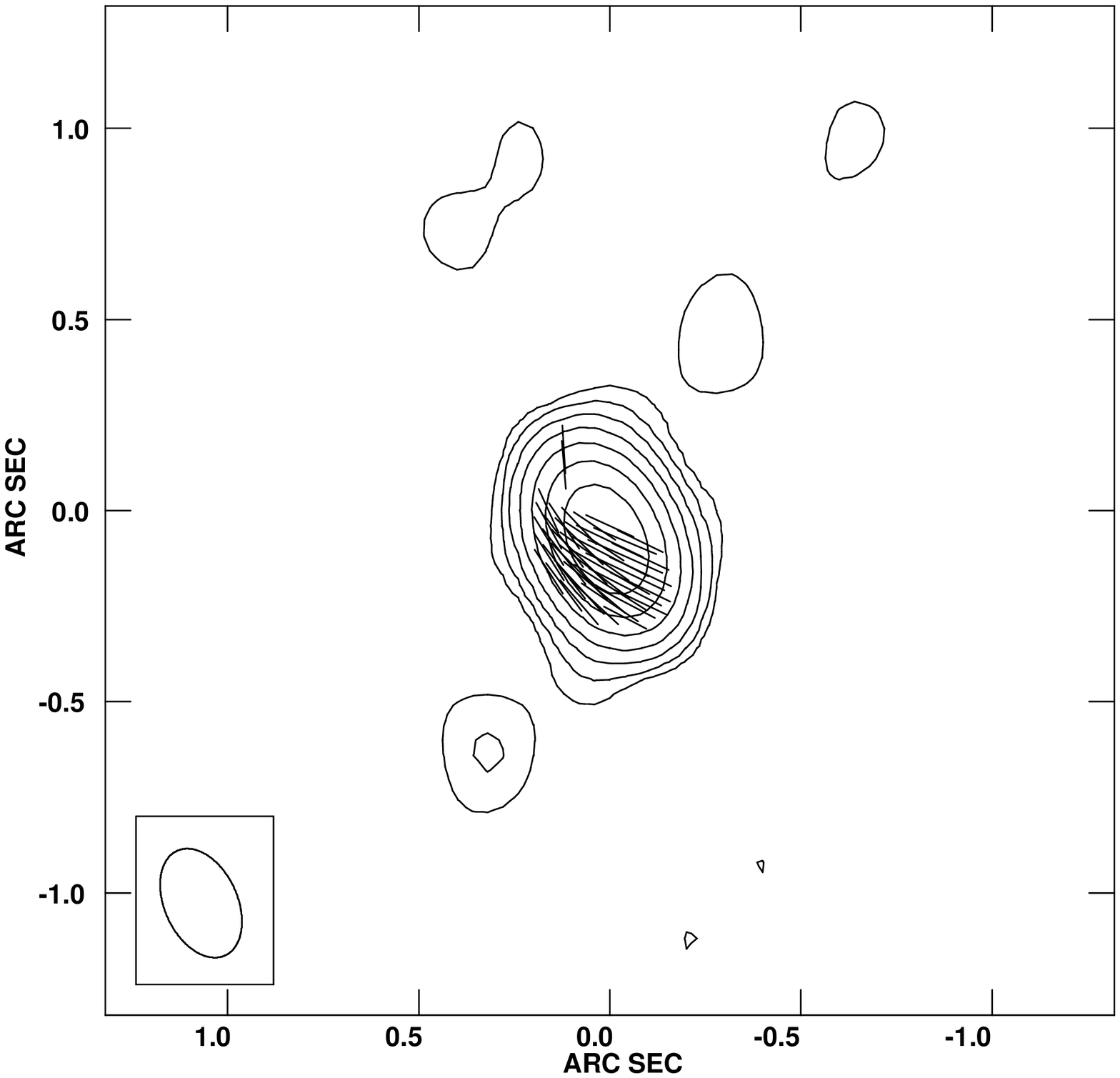}
       \label{fig:third_sub}
    }
    \subfigure[2006 December 27]
    {
\includegraphics[width=6.7cm, angle=-90]{06DEC27.ps}
        \label{fig:first_sub}
    }
\caption[sticks]{\label{maps-B}MERLIN images of GRS\,1915+105 at 1.6~GHz (top left to bottom right) centred on Right Ascension~19~15~11.54801 Declination~+10~56~44.7662, on 2003 March 24, April 18, May 09, June 15, 2006 December 24 and 27. The restoring beam is shown in the lower corner of each image and the stick vectors represent the mean position angle of linear polarisation with an intensity of 12.5 mJy per arcsecond. The contours are $3\sigma\times-1, 1, 2, 4, 8, 16....n$. (a) - 2003 March 25. $3\sigma=$~0.422 mJy with a peak flux density = 110 mJy. (b) - 2003 April 18. $3\sigma=$~0.822~mJy with a peak flux density = 86.5~mJy. (c) - 2003 May 09. $3\sigma=$~0.48~mJy with a peak flux density = 43.0~mJy. (d) - 2003 June 15. $3\sigma=$~1.18~mJy with a peak flux density = 49.7~mJy. (e) - 2006 December 24. $3\sigma=$~1.14 with a peak flux density = 144.5~mJy. (f) - 2006 December 27. $3\sigma=$~0.69~mJy with a peak flux density = 17.7~mJy.}
\end{figure*}

\begin{figure*}
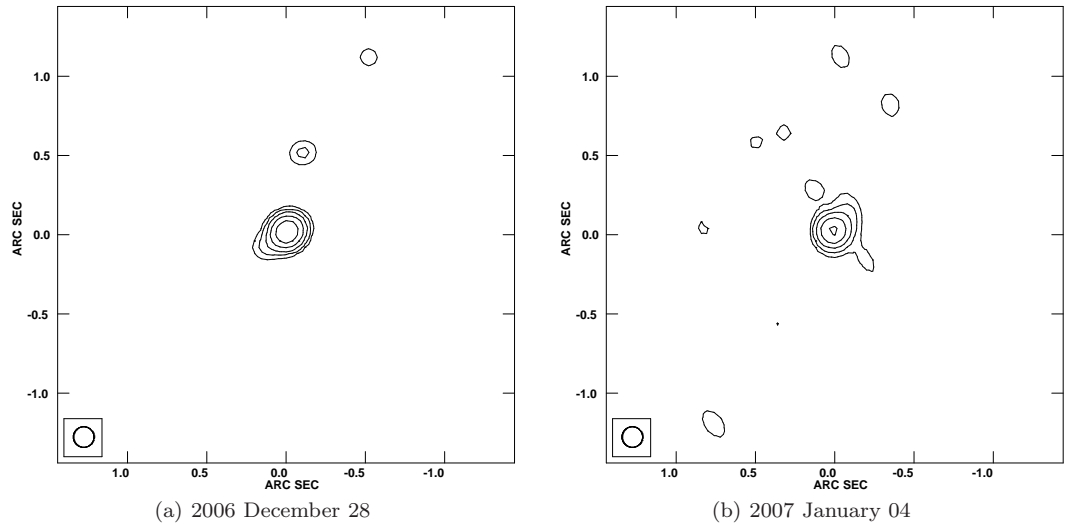

    \centering
    \subfigure[2006 December 28]
    {
\includegraphics[width=6.7cm, angle=-90]{06DEC28-nofooter.ps}
        \label{fig:second_sub}
    }
    \subfigure[2007 January 04]
    {
 \includegraphics[width=6.7cm, angle=-90]{07DEC04-nofooter.ps}
        \label{fig:third_sub}
    }
   
\caption[sticks]{\label{maps-C}MERLIN images of GRS\,1915+105 at 1.6~GHz centred on Right Ascension~19~15~11.54801 Declination~+10~56~44.7662,on 2006 December 28 and 2007 January 04. The restoring beam is shown in the lower corner of each image and the stick vectors represent the mean position angle of linear polarisation with an intensity of 12.5 mJy per arcsecond. The contours are $3\sigma\times-1, 1, 2, 4, 8, 16....n$. (a) - 2006 December 28. $3\sigma=$~0.64~mJy with a peak flux density = 19.8~mJy. (b) - 2007 January 04. $3\sigma=$~0.78~mJy with a peak flux density = 14.9~mJy.}
\end{figure*}

\section{Results and polarisation measurements}
\label{MERLIN_results}

The target source was detected in all 14 MERLIN observations, with a total flux density varying between 20 and 300~mJy at 1.4 or 1.6~GHz. Most epochs revealed a totally unresolved core, except six which showed a slightly extended structure with a position angle to the south-east. Linear polarisation was found with MERLIN in 10 out of the 14 epochs and at both frequencies of the VLBA observation. Individual 1~MHz MERLIN channels were able to detect the linear polarisation with a position angle accuracy of $\pm10\degree$; however, no significant change in the position angle of the polarisation was found across either bandpass (i.e. between $1401-1415$~MHz or $1651-1665$~MHz). Likewise a clear change in the position angle between 1.4 and 1.6~GHz was inconclusive due to high variability and low image fidelity.

Continual monitoring with the RT has shown frequent, variable radio flaring over the last decade. Strong radio flares ($>100$~mJy) are associated with a hardening of the X-ray emission (HR2~$>1.4$) and a persistently soft flux ($2-12$~keV) of $\approx50$~counts/second with a low RMS. This state has previously been identified as the `plateau' state by~\cite{1999ApJ...519L.165F}. Weaker radio flares ($\ll100$~mJy) do not clearly couple to the X-ray emission, suggesting an underlying mechanism not related to X-ray state changes.

\subsection*{Individual events}

\subsubsection*{October 1997 (MJD 50730-50750)}

In October 1997 the RT observed the first period of a sustained radio outburst, begining around 1997 October 04 (MJD~50725) and lasting approximately 30 days, ending with a short flare peaking at 220~mJy (at 15~GHz). A series of 5~GHz MERLIN observations were reported by \citep{1999ApJ...519L.165F}, showing the first Galactic superluminal proper motion with MERLIN. \cite{2000ApJ...543..373D} also reported near simultaneous VLBA observations, showing an extended jet to the south-east (SE). Our analysis of the VLBA data has detected linearly polarised emission from within the extended jet at $2.5\%$ and $7.2\%$ fraction of the total intensity (I$_{\rm{total}}$) at frequencies of 2.27~GHz and 8.3~GHz, respectively, but none within the core.

Over this period the RXTE-ASM detected a sudden quenching of the soft X-rays, with a sharp rise in the hardness ratio, HR2~$> 1.6$. The ASM count rate then remained at a level of $\approx50$~counts/second, with a low RMS fluctuation. This state ended with a transition to a different state with a short radio and soft X-ray flare of~$\approx150$~counts/second, associated with the ejection of a superluminal knot reported in \cite{1999ApJ...519L.165F}.

\subsubsection*{April - July 1998 (MJD 50905-51020)}

A major radio outburst was detected by the RT starting around 1998 April 04 (MJD~50907), which lasted for~$\approx~110$~days. During this period the RT also detected multiple short flares, with the brightest flare reaching 362~mJy (at 15~GHz). The X-ray emission returned to the plateau state of $\approx50$~counts/second, as seen in October 1997, and was notably harder during this outburst, with the ASM hardness ratio reaching HR2~$\sim2.0$. The multiple short radio flares can also be seen to be associated with the soft X-ray flares throughout this quenched state (as shown in Figure~\ref{monitor_1997-1998}).

A MERLIN observation on 1998 June 09 (MJD~50973) was made six days after one of the short flares~(Figure~\ref{maps-A}-a),\linebreak showing an extended jet to the SE of the core. A total flux of 145~mJy (at 1.6~GHz) was detected compared to a RT observation of 79~mJy (at 15~GHz), giving a spectral index of $\alpha=-0.3$ (where $S \propto \nu^\alpha$). Linear polarisation was detected by MERLIN at a total flux of 9.5~mJy, which was $\sim7\%$ of~$I_{\rm{total}}$.

\subsubsection*{April 1999 (MJD 51270)}

Paradoxically, the strongest radio flare detected in these MERLIN observations occured during a period of relative inactivity in radio and X-rays. On 1999 April 02 (MJD~51270) MERLIN 1.6~GHz observations detected a peak flux density of nearly $\approx300$~mJy, which quickly decayed to $\approx170$~mJy. Little evidence of a jet was found in the image~(Figure~\ref{maps-A}-b) with only a marginally resolved core detected. Linearly polarised emission was detected in the core at a total flux of 3.5~mJy, which was $\sim1.5\%$ of $I_{\rm{total}}$.

The RT did not detect a sustained period of radio outburst and only measured a short flare with a relatively modest flux density of $\approx60$~mJy (at 15~GHz), giving an approximate spectral index of $\alpha \sim -0.6$. RXTE-ASM observations during April 1999 also had a relatively low count rate of $45-60$~counts/second, showing no sign of an X-ray flare during this period.

\subsubsection*{November - December 1999 (MJD 51500-51550)}

The latter two months of 1999 had a modest period of radio activity following a period of relative quiet. The RT measured short flares which peaked at 126~mJy and 171~mJy (at 15~GHz), and a sustained period of radio out-bursting occurred for $\sim15$~days. Although continual RXTE-ASM observations were not available, X-ray counts peaked at~$\approx~210$~counts/second and the hardness radio reached HR2~$\sim2$. 

MERLIN observations on 1999 November 18 (MJD~51500) detected another short, strong radio flare of 221~mJy (at 1.6 GHz), whilst the RT measured $\sim90$~mJy (at 15~GHz), making a spectral index of $\alpha=-0.4$. Again only a marginally resolved core was found~(Figure~\ref{maps-A}-c), with linearly polarised emission at a fractional level of $\sim3.2\%$. Four days later on 1999 November 22 (MJD~51504) MERLIN observed that the radio flare had quickly dissipated to 37~mJy~(Figure~\ref{maps-A}-d), showing no linearly polarised emission ($\leq0.4\%$). At 15~GHz there was a similar flux density of 30~mJy, resulting in dramatic change in the spectral index, which reduced to $\alpha=-0.2$.

Finally on 1999 December 28 (MJD~51540) a small flare of 90~mJy was detected with MERLIN at 1.6 GHz and RT at 15~GHz with 40~mJy ($\alpha \sim -0.4$). A very unusual structure was seen to north of the core~(Figure~\ref{maps-A}-e), not normally associated with the position angle of previous jets. The linear polarisation was also found to be unusually high for a frequency of 1.6~GHz with polarised flux of~12.7~mJy, i.e. $\approx14\%$ of $I_{\rm{total}}$.

\subsubsection*{February - July 2003 (MJD 52695-52825)}

The most spectacular event occured in the first half of 2003, in a giant outburst of radio activity, with multiple short radio and X-ray flares. The event began on 2003 March 02 (MJD~52700) with the RT observing a short flare at 15~GHz of 151~mJy, followed 14 days later with peak flare of 372~mJy and then a slow decay over the next $\sim100$~days. The RXTE-ASM, as seen in 1997, observed a quenching of the soft X-rays, with a strong rise in the hardness ratio and the emission remaining at a level of $\approx50$~counts/second, with short X-ray flares occuring with radio flares.

Five MERLIN observations were taken over this period of activity [starting four days after the initial radio flare on 2003 March 06 (MJD~52704)]. This epoch showed only a slight NW-SE extension~(Figure~\ref{maps-A}-f), suggesting a small jet associated with the initial outburst with a weak flux density of $\approx31$~mJy (at 1.6~GHz), whilst the RT measured 15~mJy (at 15~GHz), giving a spectral index of $\alpha=-0.4$. Only weak linearly polarised emission was found near the small jet, at a fraction of the total emission of $5.2\%$.

The second observation occured on 2003 March 25 (MJD 52723), seven days after the peak flare was detected with the RT. The MERLIN map showed very little extension~(Figure~\ref{maps-B}-a), but had a variable core of $\approx 120$~mJy (RT data was not available). Polarised emission was again detected within the core at a fractional level of $\approx 5.6~\%$. No X-ray flares were detected with the ASM instrument over this time.

On 2003 March 05 (MJD 52734), 35 days after the initial radio flare, the RXTE detected a short, but relatively strong and soft X-ray flare of 150 counts/second. This was followed by another increase in radio emission and then a return to the constant hard X-ray state. MERLIN~(1.6~GHz) observations 13 days after this short event, on 2003 April 18 (MJD~52747), revealed an extended component of 20~mJy to the south-east, with the core at a peak flux density of~135~mJy~(Figure~\ref{maps-B}-b). Measurements at 15 GHz with the RT showed a strong flare of 143 mJy, suggesting the core had an approximately flat spectrum. Linearly polarised emission was detected both within the core (2.4~mJy) and the jet (1.3~mJy) at a fractional total emission of $2.6\%$ and $6.5\%$ respectively. On 2003 May 09 (MJD~52768), 34 days after the strong X-ray flare, MERLIN observations showed the extended emission had disappeared~(Figure~\ref{maps-B}-c) and an unresolved core had decayed to $43$~mJy at 1.6~GHz with linear polarisation at $6.3\%$ (no RT data were available).

The 2003 `plateau' state ended on 2003 May 29 (MJD~52788) with an X-ray flare of 163 counts/second and the start of a new period of highly variable soft emission. Shortly following this state change, MERLIN observations were taken on 2003 June 15 (MJD~52805).  The image showed an extended jet with a total flux of 67~mJy, which was now stronger than the core which had a peak flux 41~mJy~(Figure~\ref{maps-B}-d). Furthermore, the linearly polarised emission was no longer detectable from within the core and could only be detected within the jet (RT was also not available). 

\subsubsection*{December 2006 - January 2007 (MJD 54091-54105)}

RATAN observations on 2006 December 22 (MJD~54091) detected a flare of 350~mJy at 1.0 GHz, 150~mJy at 2 -- 8 GHz and 80~mJy at 11.2~GHz (i.e. a steep spectrum), which triggered the final four MERLIN epochs in December 2006 and January 2007. Observations were taken simultaneously at 1.4 and 1.6~GHz by switching between frequencies every hour; however, only images produced at 1.6~GHz are shown in Figures~\ref{maps-B} and \ref{maps-C} as there were no structal differences between respective frequencies (i.e. only the peak density flux changed).

The MERLIN observations on 2006 December 24 (MJD~54093), confirmed the bright flare of $\sim145$~mJy and $\sim157$~mJy at 1.4~GHz and 1.6~GHz respectively~(Figure~\ref{maps-B}-e) . An unresolved core showed linear polarisation associated at a fractional level of $5.3\%$ and  $7.3\%$ at 1.4~GHz and 1.6~GHz respectively. The RATAN spectrum showed that the spectrum had flattened, suggesting the optical depth had increased.

However, this was only a short lived flare as on 2006 December 27 (MJD~54096), just five days after the initial detection, the flare had decayed to 21~mJy at 1.4~GHz and 27~mJy at 1.6~GHz~(Figure~\ref{maps-B}-f), with RATAN also showing the source to be much weaker. A slight extension to the SE was detected, suggesting the formation of a weak jet. The source remained in a similar weak state for the remaining two epochs, on 2006 December 28 (MJD~54097) and 2007 January 04 (MJD~54104)  and all linear emission had reduced to~$<0.5\%$ (see Figures~\ref{maps-C}-a and~\ref{maps-C}-b).

RXTE observations (Figure~\ref{monitor_1999-2003-2006}~bottom) showed no obvious spectral transition and the radio flare was not associated with an X-ray flare.

\section{Discussion}

This long-term comparative study between the X-ray and radio emission from GRS\,1915+105 shows a complex and varied display of the emission mechanisms in a relatively bright X-ray binary. When studied over a ten year period, characteristics in both the radio and X-ray have repeated in quasi-periodic patterns. 

Broadly summarizing, the observed radio emission is attributed to relativistic jets, which are coupled to the accretion disk in a poorly understood process. Emission from the accretion disk is dominated by X-ray emissions that can be modeled by a multi-temperature blackbody and a high-energy, non-thermal component. This non-thermal component, fitted with a power-law, is very likely to be due to the comptonization of cold electrons surrounding the jet and accretion disk. 

Understanding the interplay between the radio jet, thermal X-rays and the non-thermal component, requires the classification of the different `states' as described in \cite{1996ApJ...467L..81F}. Monitoring with the RT and the RXTE-ASM, has therefore been used to classify the state of GRS\,1915+105 into the following: Weak radio emission ($\ll100$~mJy) with short-lived flares that do not clearly couple to the X-ray emission - this has been identified as the so-called `radio-quiet state'; strong radio outbursts ($>100$~mJy), typically lasting for weeks, that are associated with a hardening of the X-ray emission and a persistently soft flux - this has been identified as the so-called `plateau state'; transitions from the `plateau' state accompanied by a short X-ray flare and the ejection of superluminal knots - this has been identified as the so-called `flaring state'.

\subsection{No detected jet - `Radio quiet state/Short flares'}

The behaviour of the jet is least understood in the radio quiet state of GRS\,1915+105. The radio emission is characterized by a weak flux that is typically less than 5~mJy at 15~GHz. Short radio flares are occasionally interleaved, which last less than a day and can reach a peak flux of $100-200$~mJy at 15~GHz. X-ray emission shows no clear correlation with the radio during this state and typical RXTE-ASM peaks of $100-150$~counts/second are observed. 

Due to selection effects that limit most observations to periods of strong radio emission, only a few high-resolution observations have been taken in this state. e-EVN observations taken in this state during April 2006~\citep{2007MNRAS.374L..47R} showed no evidence for any compact jet, unlike the jet observed by \cite{2003A&A...409L..35F}, suggesting no outflow of material. Likewise, MERLIN observations taken in 1999 showed no large scale structure, despite X-rays flares, suggesting a de-coupling of the accretion disk to any radio emission. 

MERLIN observations taken at the end of 2006, observed a short radio flare of $\sim150$~mJy at 1.6 GHz. This emission quickly decayed to  less than 20 mJy, again showing no large scale ejections. RATAN observations also followed the short flare and measured an optically thin spectral index that quickly flattened as the flare decayed. The change in spectral index indicates that material did flow out of the X-ray binary; however, no superluminal knots were formed and nor was a steady jet. This is possibly either due to the jet not becoming energetic enough or interactions with the stellar system  preventing the release of material.

\subsection{Steady compact jet - `Plateau state'}
\label{sec:steady-jet}

The plateau state in GRS\,1915+105 is analogous to the low/hard state of most XRBs, like A0620-00, as the X-ray spectrum becomes slightly harder and is associated with strong radio emission. \cite{2000A&A...355..271B} identified these X-ray states as class~$\chi$ using the RXTE-PCA. However, this analogy (with the low/hard state) is not entirely accurate as the X-ray spectra still remains relatively soft compared to other low/hard state XRBs~\citep{2006csxs.book..157M}. Physically this means that the X-ray emission still has a significant contribution from the thermal accretion disk, whilst producing  steady compact radio jets. As such, this accretion state in GRS\,1915+105 shows unique coupling between the in-falling material and the ejected jets.

All long periods of radio outburst (typically lasting $\sim100$~days) are associated with the plateau state. VLBI observations taken by \cite{2003A&A...409L..35F} in April 2003 and similarly in \cite{2000ApJ...543..373D} showed a steady, compact jet on AU-scales. Radio emission was optically thick as the jet was partially self-absorbed. This radio flux appeared to have a close relationship to the X-rays, which also emitted at a steady rate throughout this state. MERLIN observations taken in 1998 and over the April 2003 outburst, showed no large scale ejections, unless there was an X-ray flare; therefore, this radio emission is constrained relatively close to the accretion disk and this is the best state to study the coupling between the radio and X-ray emission.

The cause of the emission in the plateau state is possibly due the collapse of an inner accretion disk through instabilities in the accretion process. This would cause the accretion to be dominated by an advection-dominated process (e.g. ADAF), rather than a standard accretion disk. Advection-flow accretion also causes the comptonization of a surrounding corona of cold electrons, hence the hardening of the X-ray spectra from a more dominant non-thermal X-ray component. The inner disk would then be re-filled from a surrounding outer disk, and form the compact jet.

\subsection{Ejection of knots - `Flaring state'}
\label{sec:knots}

It has been well established that a state transition in GRS\,1915+105 is associated with the ejection of superluminal knots~\citep{1999MNRAS.304..865F}. However, the precise progenitor of the knot has not been clearly identified nor has the exact cause of their formation. \cite{2005MNRAS.363..867M} suggested that an increase in the jet velocity causes internal shocks to form and produce the ejected knots.

\subsubsection*{Proper motion of knots}

\begin{table*}
\centering
\begin{tabular}{|ccccc|}
\hline
\rowcolor[rgb]{0.8,0.8,0.8} MERLIN epoch & Separation & X-ray flare & Proper motion & Intrinsic velocity \\
\rowcolor[rgb]{0.8,0.8,0.8} (MJD) & (mas) & (MJD) & (mas d$^{-1}$) & (c) \\
\hline
1998 June 09 (MJD~50973) & $165\pm15$ & 50967 & $27\pm3$ & $1.1\pm0.2$ \\
\hline
2003 March 06 (MJD~52704) & $180\pm20$ & 52694 & $18\pm2$ & $0.79\pm0.13$ \\
\hline
2003 April 18 (MJD~52747) & $235\pm5$ & 52733 & $16.5\pm1$ & $0.74\pm0.063$ \\
\hline
2003 June 15 (MJD~52805) & $305\pm5$ & 52788 & $17.5\pm1$ & $0.77\pm0.063$ \\
& & 52792 & $23.5\pm1$ & $0.97\pm0.07$\\
\hline
\end{tabular}
\caption[List of published proper motion of GRS\,1915+105 before 2006]{The proper motions of different ejecta from GRS\,1915+105, based on the time after a short X-ray flare. The intrinsic velocity is modelled assuming a distance of $12.5\pm1.5$~kpc and $\beta\cos\theta=0.41$.}
\label{table:knot_motion}
\end{table*}

By identifying the precise times of the formation of knots, it is then possible to derive their proper motions. This assumes no deceleration of the jet, as supported by \cite{2007MNRAS.375.1087M} who found no deceleration beyond  70~mas from the core. The proper motion of an approaching, $\mu_{app}$, or receding, $\mu_{rec}$, plasmon is given as

\begin{equation}
\mu_{app,rec}=\frac{\beta \sin \theta}{(1\mp\beta \cos \theta)} \frac{c}{D},
\label{eq:velocity}
\end{equation}

\noindent where $\theta$ is the jet angle to the line of sight, $\beta$ ($=v/c$) the intrinsic jet velocity and D the distance to the source. The quantity $\beta\cos\theta$ can be measured independently of distance from previous observations using

\begin{equation}
\beta \cos \theta = \frac{\mu_{app} - \mu_{rec} }{\mu_{app} + \mu_{rec}},
\label{eq:angle}
\end{equation}

\noindent where both the approaching and receding components were observed. MERLIN observations by \cite{1999ApJ...519L.165F} found $\beta\cos\theta=0.41\pm0.02$ and VLA observations by \cite{1994Natur.371...46M} found $\beta\cos\theta=0.323\pm0.011$.

Then, by using Equations \ref{eq:velocity} and \ref{eq:angle} we can find the intrinsic velocity of the jet

\begin{equation}
\beta=\sqrt{ \frac{\mu^2_{app}D^2(1-\beta \cos \theta)^2}{c^2}+(\beta \cos \theta)^2},
\end{equation}

\noindent 
assuming there is no precession of the jet (i.e. $\beta \cos \theta$ is a constant). The MERLIN results presented in Section~\ref{MERLIN_results} showed four epochs with extended knots that could be identified with X-ray flares. Table~\ref{table:knot_motion} shows the derived proper motions and modelled intrinsic velocities. Assuming a distance of $12.5\pm1.5$~kpc from HI absorption measurements, these results provide evidence that the velocity of the ejected components was intrinsically different for each event.

In 1998 June 09 (MJD~50973) a very high proper motion was observed of 27 mas d$^{-1}$ [c.f. VLA $\mu_{app}\sim17.6\pm0.4$~mas \citep{1994Natur.371...46M}; MERLIN $\mu_{app}\sim23.6\pm0.5$~mas \citep{1999ApJ...519L.165F}]; this implies an apparent motion of $1.5~c$ (again assuming a distance of $12.5\pm1.5$~kpc) and an intrinsic velocity of $1.1\pm0.2~c$. This large velocity potentially violates the speed of light and can be explained as either (i) the incorrect identification of the ejection date (i.e. not coincident with the X-ray flare), (ii) the jet had a much higher value for $\beta\cos\theta$ or (iii) an incorrect distance to the source has been estimated. As only one X-ray flare occurred during this period, it is unlikely that (i) can explain the high velocity. Also, as the same distance was used for all four knots (iii) does not explain the significantly higher velocity found in the first knot. Therefore, this effect is most likely due to (ii) - a change of the jet angle to the line of sight.

Table~\ref{table:knot_motion} shows that the ejections in 2003 exhibited a much lower proper motion of approximately 18 mas d$^{-1}$, as described by \cite{1999MNRAS.304..865F} and \cite{1994Natur.371...46M}. The initial observation on 2003 March 06 revealed only a small extension associated with the start of an active period. Following this, the VLBA observation by \cite{2003A&A...409L..35F} found evidence of only a highly compact jet on 2003~April~02~(MJD~52731), shortly before an X-ray flare, but MERLIN observations on 2003~April~18~(MJD~52747), 13 days later, revealed the ejection of a plasmon clearly associated with this event. The solitary X-ray flare on 2003~April~05~(MJD~52734) is therefore the most probable progenitor of the observed plasmon and would make its proper motion $16.5\pm1$~mas~d$^{-1}$. Interestingly, GRS\,1915+105 returned to the plateau state demonstrating that only a short ($\sim$~day) change in its X-ray state is required to eject material and the underlying emission from the plateau state had remained. This supports the idea that the knots were formed from internal shocks, rather than a cataclysmic destruction of the jet-disk model. It also demonstrated that the knots were formed soon after the X-ray spectral change.

The period of activity in 2003 finally ended in a similar manner to the end of the 1997 event. MERLIN observations on 2003 June 15 (MJD 52805) showed another ejection of a possible superluminal plasmon. As this event ended with multiple X-ray flares an exact proper motion cannot be identified from the MERLIN observations.

\subsection{Variations in polarisation}

GRS\,1915+105 shows a large variation in the fraction of linear polarisation in the core from less than $0.2~\%$ to about $14~\%$. During the ejection of material, the extended jet shows polarised components whilst the core appears depolarised. The electric field vectors of the jets are orientated along the jet axis if the rotation measure is low and the magnetic field is therefore perpendicular to the jet. This is suggestive of a tangled magnetic field compressed by a shock front travelling along the jet.

The low levels of polarisation in the core can be explained if the inner emitting jet passes through various depths of surrounding Faraday material. Differential degrees of Faraday rotation would cause the cancellation of linear polarisation, until the emitting material is far enough from the central region.

As recently ejected components are likely to be in a much denser surrounding medium than those that were ejected earlier and have moved away~ \citep{2007MNRAS.375.1087M}, one would expect the Faraday depth to fall as the components moves outward and depolarisation would be expected to decrease and the fractional polarisation would increase. This is in contrast to \cite{1999MNRAS.304..865F} who found the fraction of linear polarisation to quickly decay from $\sim13\%$ to $\sim6\%$ in the approaching jet. Additionally, \cite{1995ApJS..101..173R} reported no Faraday rotation in earlier VLA observations, showing no changing in position angle between 5.0, 8.4 and 15 GHz, resulting in a RM$<50$~rad~m$^{-2}$. Observing no change in position angle across the MERLIN sub-band is thus consistent with the VLA results and constrains RM$<600$~rad~m$^{-2}$ (at a $3\sigma$ limit). 

The multi-wavelength VLBA observations in 1997 were taken after a state change (from the plateau state) associated with superluminal knots. Linearly polarised emissions were detected from the extended jets; however, no linear polarisation was detected from the core. The results presented in Section~\ref{MERLIN_results} showed the linear polarisation was a higher fraction of the total jet emission at the higher frequency. Whilst the lack of linear polarisation from the core can be explained by intrinsic changes in the outflow (i.e. simply the absence of linearly polarised emission), this cannot account for the changes with respect to frequency; changing levels of fractional polarisation with frequency is predicted by the presence of a Faraday rotating medium. These arguments therefore support the observed variation in  polarisation as due to partial depolarisation by Faraday rotation.

\section{Conclusions}

GRS\,1915+105 has been simultaneously observed in the radio and X-rays in various different states over a ten year period. This comparative study between the X-ray and radio variability shows, for the first time, both long term ($\sim$~weeks/months) and short lived (intra-day) variations. Daily monitoring shows periods of extended activity and X-ray spectral states that are related to the ejection of material. Radio activity over this period has shown to originate from the central 150 mas core, unless a short X-ray flare is observed that marks the ejection of a superluminal plasmon. 

The proper motion of ejecta has been calculated by identifying the time of ejection and exhibit no deceleration. Measured velocities were in agreement with previous observations, and show a significant variation in velocity. Polarisation measurements show strong linear polarisation variation. Depolarisation in the core of the XRB is found during the ejection of a plasmon and the multi-wavelength observations show an increase of linear polarisation with frequency, suggesting internal Faraday rotation effects.

Observing GRS\,1915+105 over a long period of time has shown the relativistic out-flow in different forms. If the accretion conditions permit, collimated and steady jets persist over many weeks. Different apparent  velocities of superluminal knots suggest either a precession of the jet angle or variation in their formation. Finally, the nature of the short-lived radio flares remains unknown; for example, it is difficult to explain why these bright, but short-lived events do not produce extended structure, in contrast to a flare during a state transition.

\section{Acknowledgements}

The authors thank Anita Richards, Vivek Dhawan and Bill Cotton for their invaluable input to this work. Ian~K.~Brown contributed to the calibration of the MERLIN polarisation. AR acknowledges support from an STFC studentship during this research and part of this work was also supported by the EU EXPReS project. EXPReS is an Integrated Infrastructure Initiative (I3), funded under the European Commission's Sixth Framework Programme (FP6), contract number 026642. The X-ray data was provided by the ASM/RXTE teams at MIT and at the RXTE SOF and GOF at NASA's GSFC. MERLIN is a national facility operated by the University of Manchester and partially supported by STFC. The Ryle Telescope is operated by the University of Cambridge and supported by STFC.

\bibliographystyle{mn2e}
\bibliography{rushton}

\label{lastpage}

\end{document}